# Movement Sequencing: A Novel Approach to Quantifying the Building Blocks of Human Gait

Hammerberg, Alexandra G.[1], Grunblatt, Samuel K.[2], Kramer, Patricia A.[3]

## 1. Abstract

By 2050, a quarter of the US population will be over the age of 65 with greater than a 40% risk of developing life-altering neuromusculoskeletal pathologies. The potential of wearables, such as Apple AirPods and hearing aids, to provide personalized preventative and predictive health monitoring outside of the clinic is nascent, but large quantities of open-ended data that capture movement in the physical world now exist. Algorithms that leverage existing wearable technology to detect subtle changes to walking mechanics, an early indicator of neuromusculoskeletal pathology, have successfully been developed to determine population-level statistics, but individual-level variability is more difficult to parse from population-level data. Like genetic sequencing, the individual's gait pattern can be discerned by decomposing the movement signal into its fundamental features from which we can detect "mutations" or changes to the pattern that are early indicators of pathology – movement-based biomarkers. We have developed a novel approach to quantify "normal baseline movement" at an individual level, combining methods from gait laboratories with methods used to characterize stellar oscillations. We tested our approach by asking participants to complete an outdoor circuit while wearing a pair of AirPods, using orthopaedic braces to simulate pathology. We found that the novel features we propose are sensitive enough to distinguish between normal walking and brace walking at the population level and at the individual level in all sensor directions (both $p < 0.05$). We also perform principal component analysis on our population-level and individual-level models, and find significant differences between individuals as well as between the overall population model and most individuals. We also demonstrate the potential of these gait features in deep learning applications.

## 2. Introduction

From an evolutionary perspective, it is advantageous to be able to move through your environment with minimal energetic expenditure [1]. Walking mechanics are predicated on mechanical energy exchange [2,3, etc.] and this efficiency is achieved through consistent, repeating motion, with limited perturbations [4]. Also, maintaining functional sight and stability of the visual system while moving is a core requirement for survival (e.g. detection of threats, identification of food or water sources) and adaptive mechanisms for maintaining head stability across species are foundational aspects of anatomy [5,6]. In bipedal humans, the impact of ground contact during motion is consistently damped or absorbed from the feet up the body to the head [7]. This occurs through the complex mechanics of a multi-jointed musculoskeletal system (e.g., the spine with its curvature and intervertebral disks), coordinated by the nervous system (i.e., the sensorimotor pathway) and powered by the circulatory and respiratory systems. When components of this system break down (e.g., due to musculoskeletal injury, neurological degradation, cardiac and respiratory impairment), compensatory mechanisms activate to maintain functional movement. Depending on the underlying source of pathology, these mechanisms may be as simple as an overall slowing of walk velocity [8] or as complex as a change in segmental movement patterns from one step to another (i.e., decrease in gait consistency) [9,10].

---

[1] hominin.ai Inc.
[2] University of Alabama
[3] University of Washington

Walking velocity has become the sixth vital sign and other changes to gait mechanics are indicators of musculoskeletal and neurological health and overall well-being [11]. One of the core challenges in monitoring health through gait, though, is variability among individuals. While the gross pattern of bipedal locomotion is consistent across humans (i.e., propulsion and braking on alternating legs via an oscillating stance and swing phase), the details of each individual's movement patterns and mechanics are sufficiently unique to be used in identification [12]. Further, the details of individual mechanical adaptations to environmental conditions such as terrain changes (e.g., incline, decline, snow, grass) are person-specific [13]. This uniqueness presents a challenge to identifying movement-based biomarkers of pathologies or deviations from "normal" that may indicate changes to health: "Normal" is an individual baseline that is the key to personal health monitoring outside of the clinic and to early intervention.

Laboratories and specialized clinics have clearly demonstrated that camera-based imaging systems can identify biomarkers of neurological and musculoskeletal pathology with increasing precision as camera and visual processing algorithms evolve [14]. These systems, however, continue to present challenges in effective and scalable deployment for remote monitoring in the world outside of the clinic or lab. Video-based systems that capture human movement, whether they use structured, pre-labeled anatomy (e.g. physical marker-based motion capture [15]) or are pre-trained to identify points of interest on the body (e.g. automatic identification of joints [16]), rely on the quality and position of the cameras used for data acquisition and are subject to error when relevant anatomy is obscured by clothing or other objects. Specialized camera systems are costly and complicated to set up and maintain, and while research groups and companies are attempting to deploy smartphone and in-home camera-based solutions, there are still technical barriers as well as negative consumer perception to camera-based surveillance.

With the rise of movement sensor integration in consumer wearable devices such as headphones, smartphones, and watches, typically through inertial sensor units (IMUs), it is now possible to quantify a movement baseline and monitor movement biomarkers at the individual level outside of a laboratory [11]. For gait monitoring, the head is an optimal location due to its stable position in relation to the rest of the body, thus reducing signal interference, and capturing movement of the body using sensors located on the head enables us to gain a holistic view of the underlying health of the system, as described above. Wearable device manufacturers and apps that utilize their data, however, rely on movement models trained towards the population average on aggregate data [17]. There are three major restrictions to this approach: 1.) it is impossible to capture a fully representative sample from even a large participant sample, 2.) the range of variability at the population level is higher than the variability within an individual, so signal sensitivity is considerably lower in aggregate data, and 3.) the gait metrics from these models are limited in scope.

The gait metrics currently provided through commercially available wearable devices and smartphones have not progressed far beyond step count, step length, and velocity of travel, although some solutions also provide double support time, asymmetry, and company-specific scores for steadiness [17]. These metrics have not been shown to be sensitive enough biomarkers to capture early indicators of pathology in walking, although they may be useful for monitoring disease progression or injury recovery [11]. To date, the approach to gait analysis through wearables has been to reproduce metrics from laboratory data modalities like motion (position data) and force (ground reaction force data). Aside from the challenge of data scarcity (e.g. a basic full-body motion capture setup consists of ~6-13 markers distributed across the body, with position data from each marker whereas wearables are typically constrained to a single, unilateral location on the body), using IMU data to reproduce metrics derived from position and force data is a significant, unsolved challenge with an extensive literature [18] of trying to get GRF from IMUs]. Therefore, there is pressing need for an evolution in the approach we use to analyze gait using IMU data.

In this paper we leverage IMU sensors located in commercially available headphones and a novel approach to quantifying human walking. That is, we establish "normal baseline movement" for an individual by combining methodologies used in biomechanics laboratories for quantifying human gait mechanics with independently developed frequency analysis approaches used to characterize stellar oscillations. We utilize the featuresets from these methodologies to distinguish between normal walking and simulated pathology, comparing our results at both the aggregate "population" level and the level of the individual participant. We show that the signal sensitivity to pathology at the level of the individual is lost when the data are aggregated and that creating scalable individual movement models requires the development of a new approach to quantifying human gait. We also showcase the potential of these novel gait features and our individual-level modeling approach for deep learning applications by demonstrating their impact in a simple neural network. This sets the foundation for future work building multimodal models capable of learning individual human movement mechanics in a real world context.

3. **Methods**

Data from existing inertial measurement unit (IMU) sensors that are embedded in spatial audio-enabled headphones were collected through a custom mobile application, InStep Movement Analytics, that is integrated with Apple AirPods. The headphones provided raw acceleration and angular velocity data at a rate of 25Hz (+/- 5Hz). These data were accessed through Apple's open API framework, CoreMotion, which streams data from either the left or right headphone [19]. The head-based IMU tracks nine simultaneous data streams - acceleration in X, Y, and Z, rotation in X, Y, and Z, and angular position in X, Y, and Z (Figure 1). GPS location information was obtained through the Apple framework at a rate of approximately 10Hz, which was used to call data regarding environmental conditions such as weather and surface type, as well as altimeter data (collection rate 10 Hz), also through open APIs [20]. Live GPS and altimeter data were collected via the mobile application downloaded on a smartphone and worn on the body during testing.

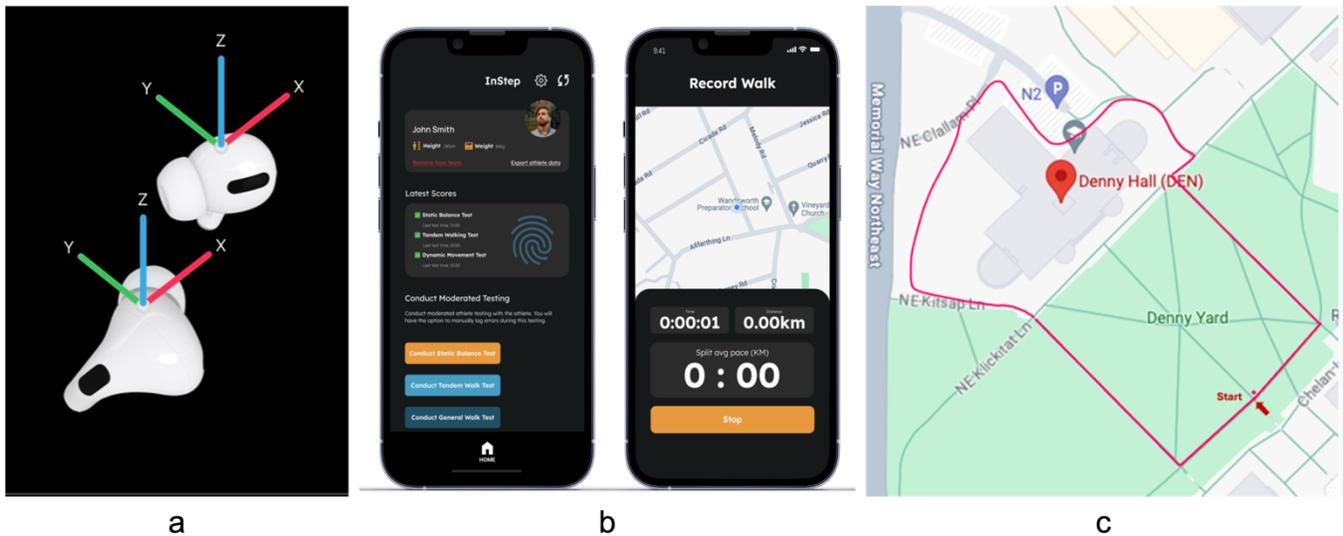

Figure 1 – a.) Positive coordinate axes for Apple AirPod Pro headphones via Apple CoreMotion [19], b.) the data collection app, c.) the outdoor circuit on the University of Washington campus that participants were asked to walk.

a. Human Subjects Protocol

This study simulated musculoskeletal impairments / pathologies in the lower limb and lumbar spine using orthopaedic braces. The aim was to assess the feasibility of measuring changes to individuals' gait mechanics using the IMU data from AirPods. As part of this study, participants without known pathologies were asked to complete an outdoor circuit (~0.48km) on sidewalks around a campus building at the University of Washington while wearing the headphones (Figure 1); first walking normally to establish an individual baseline, then repeating the same circuit while wearing, in turn, a stiff ankle brace, a hard knee brace, and a lumbar back brace (order randomized). On average, the outdoor circuit took 5 minutes and 9 seconds to complete during normal walking and 5 minutes and 17 seconds to complete while walking with a brace (across all brace types).

Data were collected from a convenience sample of 17 participants (11 female, 6 male), with an average age of 33 years old (range: 18 - 67 years old). Participant information is included in Table 1. Prior to data collection we collected demographic and basic anthropometric data from each participant, and asked participants about prior or current musculoskeletal pathologies. In our sample, 11 participants responded that they had experienced prior musculoskeletal pathology, none of the participants were actively experiencing musculoskeletal pain that prevented them from completing the protocol. Participant 003A's knee brace test was removed from the dataset as the brace slid off the participant's leg during the trial.

| Participant ID | Age (years) | Sex | Height (cm) | Weight (kg) | Prior MSK Pathology |
|---|---|---|---|---|---|
| 101A | 30 | M | 198 | 102 | NO |
| 001A | 18 | M | 186.7 | 73 | NO |
| 002A | 19 | F | 157.5 | 68.6 | YES |
| 003A | 23 | F | 167.6 | 82.8 | YES |
| 004A | 22 | F | 172.7 | 62 | NO |
| 005A | 33 | F | 179 | 71.9 | YES |
| 006A | 33 | M | 195.5 | 101.6 | YES |
| 007A | 32 | M | 178.5 | 65.7 | YES |
| 008A | 21 | F | 172.7 | 82.6 | YES |
| 009A | 20 | F | 163 | 86 | YES |
| 010A | 49 | F | 173.5 | 95.5 | YES |
| 011A | 21 | F | 162.5 | 58.6 | NO |
| 012A | 42 | F | 173 | 64.1 | NO |
| 013A | 35 | M | 180 | 106.2 | YES |
| 014A | 33 | F | 177.5 | 95.2 | YES |
| 015A | 62 | F | 169.5 | 94 | YES |
| 016A | 67 | M | 178 | 96 | YES |
| Average | 32.9 | | 175.6 | 82.7 | |
| StDev | 14.7 | | 10.8 | 15.8 | |

Table 1 - Participant information.

b.  Feature Extraction

We sliced the data at regular time intervals (~50 seconds), then extracted two movement featuresets from each slice of the raw acceleration and rotational signal recorded by the headphones during wear, resulting in ~1000 measurements per slice for each acceleration and rotation direction. In this initial work we focused on acceleration in vertical, mediolateral, and anteroposterior (Z, X, and Y in the headphone coordinate space), and mediolateral rotation (Y in the headphone coordinate space). Unperturbed human bipedal locomotion (i.e. walking, running, etc. without obstacles) can be characterized as (quasi)periodic, and, thus, general movement features can be characterized from the frequency domain. As described above, motion signals at the head are especially clear, minimizing the amount of pre-processing required to extract relevant features from the signal.

We extracted our two featuresets from the frequency domain; we first decomposed the periodic motion signal into a sum of sinusoids through a Fast Fourier Transform (FFT) and identified 12 descriptive features (FFT Features, F1-F12, Table 2). From the FFT we produced feature power vectors through algorithms adapted from the motion capture literature [21]. These feature power vectors contain 6 components that describe the motion, which have been normalized by the fundamental frequency, or motion cycle (identified by max power), in each direction to remove the time component, and by power (amplitude of the fundamental frequency peak), to remove the power dependence on the number of cycles captured in the sample (Meng Features, M0-M5). This normalization process allows for a direct comparison of the spectral profile of the motion itself, removing time-based confounders such as cadence (gait cycles / minute). While Meng et.al. [21] used position trajectory data from markers placed on the body in their spectral analysis and only analyzed vertical motion, we adapted their approach to use IMU data across coordinate directions, as described above, rather than position data. These vectors describe the frequency domain signature of the motion in each direction.

In addition, we extracted 15 features of the dataset in frequency space by fitting the individual time series slices as a nonparametric Gaussian process which can be defined with a series of simple harmonic oscillator components (Grunblatt Model Features, G1-G15). Stellar oscillations can similarly be modeled in the time domain as stochastically damped simple harmonic oscillators using a linear combination of Gaussian Process kernels [22,23,24], and thus we adapted this stellar oscillation modeling approach to describe the features of human movement captured in each of our data slices. We use a combination of up to 5 simple harmonic oscillator features, each of which has three additional annotating features, as well as an additional feature describing the dataset contained in the time slice and then determine the best-fit feature values through a Hamiltonian Monte Carlo approach with 1000 model steps. We record the best-fit values and uncertainties determined by our Hamiltonian Monte Carlo analysis for each data slice.

Overall, we extracted 33 features from each slice of the raw IMU data (Table 2), in each sensor direction, resulting in 132 features per slice. All feature values were recorded in logarithmic space except the FFT features (F1-F8, below). We used a limited supervision approach to prepare the features to remove any artifacts or outliers before they were used in statistical testing and model training. Each slice of data contained the classifying label "no brace" (NB) or "brace" (B) based on the test conditions as well as a participant ID classifier.

|  | Features |
| --- | --- |
| Meng Model Features | M0 |
|  | M1 |
|  | M2 |
|  | M3 |
|  | M4 |
|  | M5 |
| Grunblatt Model Features | G1 |
|  | G2 |
|  | G3 |
|  | G4 |
|  | G5 |
|  | G6 |
|  | G7 |
|  | G8 |
|  | G9 |
|  | G10 |
|  | G11 |
|  | G12 |
|  | G13 |
|  | G14 |
|  | G15 |
| FFT Features | F1 |
|  | F2 |
|  | F3 |
|  | F4 |
|  | F5 |
|  | F6 |
|  | F7 |
|  | F8 |
|  | F9 |
|  | F10 |
|  | F11 |
|  | F12 |

Table 2 - The 33 features extracted from each slice of raw IMU data in each sensor direction.

c. Wilcox Paired Testing

We performed non-parametric Wilcoxon paired testing (built-in wilcox.test function in R) with NAs removed for each feature extracted from our data, in each sensor direction (acceleration X, Y, Z, rotation Y), to determine whether there were significant (p < 0.05) differences between the feature values of no brace (NB) and brace (B) walking. We compared NB to B values across all participants ("population") combined, then separated our data by participant and compared NB to B values for each feature in each sensor direction within each participant.

d. PCA

We then scaled and centered our data before running a Principal Component Analysis (PCA) across our combined featuresets. We used the R functions prcomp from the 'stats' package25] and factoextra [26] to run our PCA. During our data preparation, missing values were replaced with NA. As PCA requires homogeneously sized data sets, NAs were substituted with numeric zeros after our data were scaled and centered. This was done to avoid biasing the scaling and centering, while still retaining the information inherent in "missingness" (e.g., the absence of a fourth frequency peak in a data slice can be considered as informative as the presence of a fourth peak, so it is critical to mark the absence of that peak and its associated features). Similar to our Wilcox testing, we conducted PCA first across the aggregated data (population) subset by sensor direction, then ran PCA within each individual, also subset by sensor direction.

e. Neural Net

We built a simple multilayer perceptron (MLP) neural net with 6 layers and 64 nodes per layer using PyTorch. We ran the MLP over 2500 epochs, using a stochastic optimization algorithm (Adam) and a cross-entropy loss function (PyTorch), with a learning rate of 0.001. We ran each MLP train/test loop three times, resampling the data each time, and averaged the model accuracy across the three loops. This was a relatively small dataset when subset by sensor, and even smaller when subset by participant and sensor, so 80% of the data were used for training and 20% of the data were used for testing on each pass. The featuresets were used as inputs to the MLP. The model's accuracy in classifying the data NB or B was tested, first across the entire dataset, then within each individual participant. Data were subdivided by sensor (acceleration X, acceleration Y, acceleration Z, and rotation Y), as in the statistical methods above, for NB vs B classification testing.

4. **Results**

a. Wilcox Pairs Testing – Population

When aggregated across all participants ("population" level), only six features were significantly different between NB and B walking: M2 in acceleration Z, G5, G10, and G13 in acceleration Y, and G12 and F9 in rotation Y (Table 3). No features were significantly different in acceleration X (mediolateral acceleration) between NB and B walking in the population-level data. There were no features that were significant in more than one sensor direction.

|  |  |  | Population | | | |
|---|---|---|---|---|---|---|
| <0.001 *** | | | Acceleration X | Acceleration Y | Acceleration Z | Rotation Y |
| <0.01 ** | Meng Features | M0 | | | | |
| <0.05 * | | M1 | | | | |
| <0.1 . | | M2 | | | green | |
| >0.1 | | M3 | | | | |
| | | M4 | | | yellow | |
| | | M5 | | | | |
| | Grunblatt Model Features | G1 | | | | |
| | | G2 | | | | |
| | | G3 | yellow | | | |
| | | G4 | | | | |
| | | G5 | | green | | |
| | | G6 | | yellow | | |
| | | G7 | | | | |
| | | G8 | | | | |
| | | G9 | yellow | | | |
| | | G10 | | green | | |
| | | G11 | | | | |
| | | G12 | | | | green |
| | | G13 | | green | | |
| | | G14 | | | | |
| | | G15 | | | | yellow |
| | FFT Features | F1 | | | | |
| | | F2 | | | | |
| | | F3 | | | | |
| | | F4 | | | | yellow |
| | | F5 | | | | |
| | | F6 | | | | |
| | | F7 | | | | |
| | | F8 | | | | |
| | | F9 | | | | green |
| | | F10 | | | | |
| | | F11 | | | | |
| | | F12 | | | | |

Table 3 - Wilcox pairs testing between NB and B walking for each feature using the aggregated participant data ("population") by sensor direction. Only three features were significantly (p < 0.05) different between NB and B walking for the population data (denoted in green).

b. Wilcox Pairs Testing – Individuals

All participants had at least one feature that was significantly different between NB and B walking across all sensor directions (Table 4). The participant with the least difference between NB and B walking was participant 006A, who only had a difference in one feature, F6, in rotation Y. The participant with the greatest difference between NB and B walking was participant 010A, who returned 22 significant differences across all features and sensor directions. On average, participants returned 10 features that were significantly different between NB and B walking across all sensor directions. An average of 4 features were significantly different between NB and B walking for each sensor direction across all participants. Two participants (007A and 010A) had features that were significant in all sensor directions.

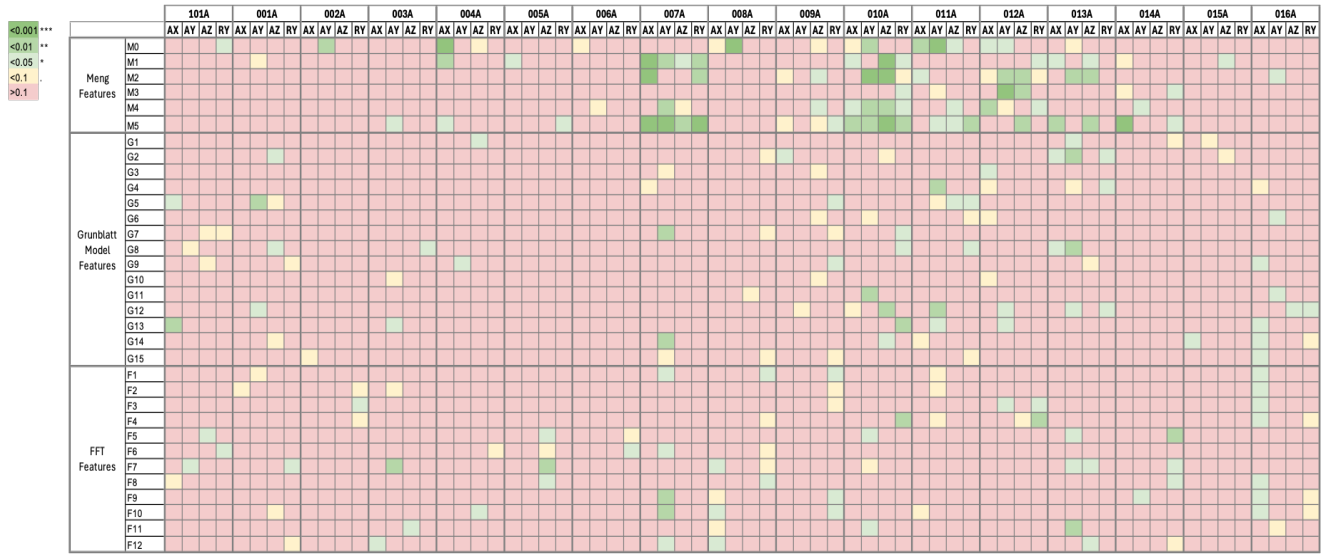

Table 4 - Wilcox pairs testing between NB and B walking for each feature, for each individual participant by sensor direction. All participants had at least one feature that was significantly different (p < 0.05) between NB and B walking. No single feature was consistently significant across all participants.

|  |  | % Participants |
|---|---|---:|
| Meng Features | M0 | 41.2% |
|  | M1 | 41.2% |
|  | M2 | 41.2% |
|  | M3 | 17.6% |
|  | M4 | 35.3% |
|  | M5 | 58.8% |
| Grunblatt Model Features | G1 | 11.8% |
|  | G2 | 17.6% |
|  | G3 | 5.9% |
|  | G4 | 11.8% |
|  | G5 | 23.5% |
|  | G6 | 5.9% |
|  | G7 | 11.8% |
|  | G8 | 29.4% |
|  | G9 | 17.6% |
|  | G10 | 0.0% |
|  | G11 | 11.8% |
|  | G12 | 35.3% |
|  | G13 | 35.3% |
|  | G14 | 23.5% |
|  | G15 | 5.9% |
| FFT Features | F1 | 23.5% |
|  | F2 | 5.9% |
|  | F3 | 17.6% |
|  | F4 | 17.6% |
|  | F5 | 29.4% |
|  | F6 | 17.6% |
|  | F7 | 41.2% |
|  | F8 | 23.5% |
|  | F9 | 23.5% |
|  | F10 | 29.4% |
|  | F11 | 17.6% |
|  | F12 | 23.5% |

Table 5 - The proportion of participants with a significant difference in the feature value between NB and B walking, across all sensor directions.

Overall, an average of 22.8% of participants (~4 participants) exhibited a significant difference between NB and B for each feature and each direction (Table 5). This varied by feature and ranged from zero participants exhibiting a significant difference between NB and B for feature G10, to 10 out of the 17 (58.8%) participants returning a significant difference between NB and B values for feature M5.

c. PCA – Population

|  |  | Acceleration X | | Acceleration Y | | Acceleration Z | | Rotation Y | | Feature Overlap |
|---|---|---|---|---|---|---|---|---|---|---|
|  |  | No Brace | Brace | No Brace | Brace | No Brace | Brace | No Brace | Brace |  |
| Population | PC1 | F10 | G15 | G5 | F9 | F10 | F10 | G15 | F9 |  |
|  | PC2 | G2 | F11 | G4 | F11 | F12 | F12 | F12 | F12 |  |
| 101A | PC1 | G9 | G14 | G9 | F11 | M0 | G14 | G7 | F12 | 1 |
|  | PC2 | G5 | G10 | G5 | F10 | G14 | M0 | F12 | G11 |  |
| 001A | PC1 | G4 | G3 | G2 | G14 | F10 | G4 | G13 | G13 | 2 |
|  | PC2 | G3 | F9 | G1 | F9 | G8 | G6 | F12 | G7 |  |
| 002A | PC1 | F11 | G1 | G10 | F9 | F9 | F11 | G14 | G4 | 2 |
|  | PC2 | G13 | F12 | G14 | F11 | F11 | M2 | G10 | G4 |  |
| 003A | PC1 | G6 | G4 | F9 | G11 | G15 | G11 | G10 | G4 | 4 |
|  | PC2 | G14 | F11 | G4 | G15 | F10 | F12 | G15 | F12 |  |
| 004A | PC1 | G4 | G15 | G13 | G4 | M3 | G15 | G4 | G5 | 3 |
|  | PC2 | G2 | G2 | G4 | F12 | G5 | G10 | G8 | G1 |  |
| 005A | PC1 | G11 | G2 | F11 | F12 | F12 | F12 | G9 | G14 | 0 |
|  | PC2 | F9 | G13 | F12 | G14 | G15 | M2 | G15 | M0 |  |
| 006A | PC1 | G1 | F9 | G5 | G5 | F11 | F12 | M0 | G15 | 1 |
|  | PC2 | G15 | G2 | M0 | F9 | G15 | F9 | G12 | G11 |  |
| 007A | PC1 | F9 | F9 | F11 | F10 | G2 | F10 | F9 | F9 | 3 |
|  | PC2 | G5 | F11 | M3 | F9 | G9 | G11 | G2 | G5 |  |
| 008A | PC1 | F10 | G5 | G2 | G15 | M0 | G1 | F12 | F11 | 2 |
|  | PC2 | F12 | G8 | G15 | M0 | F9 | G2 | F12 | G10 |  |
| 009A | PC1 | G8 | F12 | G13 | F10 | G11 | F9 | G2 | G6 | 1 |
|  | PC2 | G13 | F9 | G5 | G6 | G1 | G3 | F12 | G15 |  |
| 010A | PC1 | G8 | G13 | G13 | M0 | G1 | G6 | G1 | G5 | 0 |
|  | PC2 | G4 | G12 | G5 | G1 | G15 | G3 | G12 | F10 |  |
| 011A | PC1 | F10 | G3 | G2 | F12 | F10 | G8 | G4 | G6 | 2 |
|  | PC2 | G3 | G8 | G5 | G13 | M3 | G10 | G5 | G3 |  |
| 012A | PC1 | F10 | F11 | G2 | F11 | F10 | F9 | G13 | G13 | 3 |
|  | PC2 | G2 | G10 | G5 | F10 | G1 | G11 | G10 | G2 |  |
| 013A | PC1 | F11 | F10 | G5 | G1 | G5 | F10 | F9 | G5 | 3 |
|  | PC2 | G14 | G11 | M0 | G1 | G13 | G4 | G11 | F12 |  |
| 014A | PC1 | G4 | G5 | G8 | G15 | G7 | G11 | G3 | G5 | 1 |
|  | PC2 | G9 | G7 | G12 | G7 | F12 | G2 | G8 | G2 |  |
| 015A | PC1 | G13 | G13 | F9 | F10 | G3 | M1 | G3 | G3 | 0 |
|  | PC2 | G13 | M2 | F12 | M2 | G7 | F10 | G15 | G7 |  |
| 016A | PC1 | G2 | G3 | G3 | G14 | F10 | F10 | F12 | F10 | 5 |
|  | PC2 | G1 | F11 | G10 | F11 | F10 | F12 | G14 | F11 |  |

Table 6 - The feature with the highest importance value in Principal Component 1 (PC1) and Principal Component 2 (PC2) for Principal Component Analysis (PCA) performed across all participants (Population) and within each participant. There was no observed consistency across participants in top feature importance for PC1 or PC2. There was also little overlap between the top features in PC1 and PC2 for the population-level PCA and the participant-level PCA.

Similar to our Wilcox pairwise testing, no single feature was consistently most important to the model across all sensor directions for the population-level PCA, although F12 was the top feature of PC2 in acceleration Z and rotation Y for both NB and B walking at the population level (Table 6). At the population level, the mean relative variance explained by the top feature for PC1 was 9.1% for both NB and B walking and 13.7% and 12.5% in PC2 for NB and B, respectively, across sensor directions

(Table 7). On average, when combined PC1 and PC2 only explained 34.2% (32.0 - 37.3%) of the population-level PCA across sensor directions (Table 8).

| Test Type | PC1 Top Feature Relative % Mean | PC1 Top Feature Relative % StDev | PC2 Top Feature Relative % Mean | PC2 Top Feature Relative % StDev |
|---|---|---|---|---|
| No Brace | 9.1% | 1.0% | 13.7% | 2.0% |
| Brace | 9.1% | 0.5% | 12.5% | 3.0% |

Table 7 - The relative importance of the top feature in the population-level PCAs, averaged over all sensor directions.

| | | Variance Explained by PC1 + PC2 | | | |
|---|---|---|---|---|---|
| | | Acceleration X | Acceleration Y | Acceleration Z | Rotation Y |
| Population | No Brace | 36.1% | 35.9% | 32.0% | 37.3% |
| | Brace | 34.6% | 32.2% | 32.4% | 33.0% |

Table 8 - The total relative variance explained by PC1 + PC2 across NB and B testing for the population-level data. Combined, PC1 and PC2 explained only about a third of the relative variance in any sensor direction for both NB and B tests for the aggregated data.

d. **PCA – Individual**

No single feature was consistently most important in the PCA in any sensor direction when looking across the individual-level PCAs, although F10 appeared 15 times (22.1%) across all participants and sensor directions. Notably, there was also little overlap in the top features for PC1 and PC2 between the population-level PCAs and individuals (Table 6). Three participants (17.6%) have no overlap with the top features identified in the population-level PCA, and an additional four participants only have one top feature that overlaps with the population-level PCA, meaning that the population results are non-representative of almost half of participants.

| Participant | Test Type | PC1 Top Feature Relative % Mean | PC1 Top Feature Relative % StDev | PC2 Top Feature Relative % Mean | PC2 Top Feature Relative % StDev |
|---|---|---|---|---|---|
| 001A | Brace | 13.8% | 1.5% | 13.7% | 3.6% |
| | No Brace | 15.7% | 4.4% | 12.1% | 3.5% |
| 002A | Brace | 11.6% | 3.9% | 13.9% | 1.5% |
| | No Brace | 12.2% | 2.1% | 13.6% | 2.8% |
| 003A | Brace | 12.3% | 2.4% | 13.8% | 2.3% |
| | No Brace | 12.2% | 1.4% | 12.9% | 3.5% |
| 004A | Brace | 10.0% | 1.5% | 17.6% | 8.8% |
| | No Brace | 13.1% | 1.6% | 12.6% | 3.7% |
| 005A | Brace | 11.4% | 3.8% | 16.0% | 3.0% |
| | No Brace | 14.6% | 4.4% | 18.4% | 3.8% |
| 006A | Brace | 11.8% | 1.9% | 11.6% | 1.4% |
| | No Brace | 12.8% | 3.5% | 13.1% | 3.6% |
| 007A | Brace | 14.6% | 6.0% | 15.2% | 1.4% |
| | No Brace | 16.1% | 1.5% | 13.7% | 2.7% |
| 008A | Brace | 10.1% | 3.0% | 14.9% | 1.9% |
| | No Brace | 16.5% | 6.4% | 13.6% | 3.7% |
| 009A | Brace | 14.0% | 5.8% | 13.5% | 3.8% |
| | No Brace | 10.4% | 1.8% | 18.6% | 3.1% |
| 010A | Brace | 11.0% | 1.6% | 11.5% | 1.6% |
| | No Brace | 16.5% | 5.4% | 14.4% | 3.1% |
| 011A | Brace | 11.5% | 2.4% | 12.7% | 2.6% |
| | No Brace | 12.9% | 2.0% | 12.1% | 1.3% |
| 012A | Brace | 11.6% | 0.7% | 12.3% | 3.3% |
| | No Brace | 14.7% | 3.9% | 17.8% | 3.8% |
| 013A | Brace | 16.7% | 7.1% | 17.3% | 4.2% |
| | No Brace | 13.0% | 3.7% | 12.4% | 1.6% |
| 014A | Brace | 12.3% | 2.6% | 13.9% | 2.9% |
| | No Brace | 12.7% | 3.3% | 13.0% | 1.0% |
| 015A | Brace | 13.4% | 0.6% | 13.0% | 3.6% |
| | No Brace | 13.3% | 0.4% | 17.1% | 2.8% |
| 016A | Brace | 15.3% | 4.3% | 13.3% | 2.3% |
| | No Brace | 14.4% | 6.2% | 15.4% | 5.5% |
| 101A | Brace | 10.5% | 0.5% | 13.5% | 2.9% |
| | No Brace | 14.2% | 5.3% | 14.7% | 2.8% |
| Mean | Brace | 12.5% | 2.9% | 14.0% | 3.0% |
| | No Brace | 13.8% | 3.4% | 14.5% | 3.1% |

Table 9 - Mean relative importance of the top features in PC1 and PC2 across sensor direction by participant, averaged across sensor directions. The relative importance of the top feature is on average higher for PCAs run on each individual than for PCAs run on the population data.

On average, the relative importance of the top feature for NB walking was higher than the relative importance of the top feature for B walking (Table 9). The top features for PC1 and PC2 had consistently higher relative importance at the individual level than the top features did at population level for both PC1 and PC2. The top feature also explained more of the relative variance for individuals than it did for the population in both PC1 and PC2.

At the individual level, PC1 and PC2 explain more of the relative variance in the PCAs than they do at the population level (48.1%-76.2% across sensor directions for individual PCAs vs 32.0%-37.3% across sensor directions for population PCAs). At the individual level, the variance explained by PC1 and PC2 is consistently higher across all sensor directions in NB walking than B walking for all individuals (Table 10). There is a similar trend in the population level PCA, although the values are much closer between NB and B walking (Table 8).

|  |  | Variance Explained by PC1 + PC2 | | | |
|---|---|---|---|---|---|
|  |  | Acceleration X | Acceleration Y | Acceleration Z | Rotation Y |
| 101A | No Brace | 73.2% | 66.2% | 84.1% | 85.0% |
|  | Brace | 49.2% | 58.9% | 53.3% | 47.6% |
| 001A | No Brace | 79.0% | 78.7% | 67.8% | 80.4% |
|  | Brace | 50.2% | 50.1% | 50.5% | 45.9% |
| 002A | No Brace | 79.9% | 78.4% | 81.7% | 77.9% |
|  | Brace | 66.2% | 60.6% | 57.7% | 59.2% |
| 003A | No Brace | 73.1% | 62.1% | 54.9% | 68.8% |
|  | Brace | 46.2% | 52.7% | 50.5% | 52.5% |
| 004A | No Brace | 82.1% | 68.2% | 66.0% | 76.2% |
|  | Brace | 57.4% | 50.9% | 48.1% | 45.7% |
| 005A | No Brace | 70.1% | 68.8% | 66.6% | 72.4% |
|  | Brace | 51.6% | 52.6% | 53.4% | 52.7% |
| 006A | No Brace | 75.8% | 76.8% | 79.4% | 68.8% |
|  | Brace | 58.9% | 50.2% | 50.5% | 51.6% |
| 007A | No Brace | 81.0% | 79.0% | 70.4% | 57.9% |
|  | Brace | 56.4% | 47.1% | 48.4% | 59.7% |
| 008A | No Brace | 86.6% | 77.3% | 81.0% | 81.2% |
|  | Brace | 60.7% | 50.1% | 42.8% | 57.7% |
| 009A | No Brace | 59.2% | 61.1% | 48.6% | 66.7% |
|  | Brace | 45.0% | 44.0% | 40.0% | 43.5% |
| 010A | No Brace | 76.3% | 73.8% | 74.3% | 70.3% |
|  | Brace | 45.7% | 45.0% | 39.5% | 44.9% |
| 011A | No Brace | 70.3% | 72.8% | 72.9% | 76.0% |
|  | Brace | 56.3% | 51.1% | 41.1% | 47.8% |
| 012A | No Brace | 80.8% | 74.6% | 81.7% | 65.6% |
|  | Brace | 61.3% | 51.7% | 53.3% | 57.2% |
| 013A | No Brace | 77.1% | 60.0% | 70.8% | 74.2% |
|  | Brace | 54.0% | 47.3% | 47.9% | 49.8% |
| 014A | No Brace | 70.3% | 54.9% | 66.3% | 80.1% |
|  | Brace | 47.9% | 50.2% | 48.1% | 48.2% |
| 015A | No Brace | 83.6% | 85.3% | 76.1% | 84.4% |
|  | Brace | 47.5% | 49.7% | 53.8% | 54.2% |
| 016A | No Brace | 76.5% | 74.3% | 73.8% | 78.0% |
|  | Brace | 45.7% | 42.6% | 38.6% | 56.3% |
| No Brace | **Mean** | **76.2%** | **71.3%** | **71.6%** | **74.3%** |
|  | StDev | 6.5% | 8.2% | 9.5% | 7.3% |
| Brace | **Mean** | **53.0%** | **50.3%** | **48.1%** | **51.4%** |
|  | StDev | 6.5% | 4.6% | 5.8% | 5.3% |

Table 10 - The total relative variance explained by PC1 + PC2 across NB and B testing by participant. The average explained variance for NB was consistently higher across sensors than B.

e. **NN accuracy - Population & Individual**

The results of the MLP are presented in Table 11. The average accuracy of the MLP in identifying NB vs B walking for the population data was 63.6% (range: 61.8 - 64.6%) across all sensor directions. The average accuracy of identifying NB vs B walking by individual was 62.4% (range: 33.3 - 93.3%) across all individual and all sensor directions. There was little difference in model accuracy across sensor directions at the population level or when averaged across participants; however, the model accuracy did vary across sensor directions within participants.

|  | Acceleration X Accuracy | Acceleration Y Accuracy | Acceleration Z Accuracy | Rotation Y Accuracy |
|---|---|---|---|---|
| 101A | 66.7% | 66.7% | 50.0% | 33.3% |
| 001A | 77.8% | 66.7% | 83.3% | 66.7% |
| 002A | 50.0% | 66.7% | 41.7% | 33.3% |
| 003A | 46.7% | 46.7% | 73.3% | 75.0% |
| 004A | 40.0% | 40.0% | 66.7% | 46.7% |
| 005A | 60.0% | 46.7% | 80.0% | 53.3% |
| 006A | 53.3% | 60.0% | 40.0% | 66.7% |
| 007A | 61.1% | 77.8% | 55.6% | 66.7% |
| 008A | 93.3% | 53.3% | 33.3% | 66.7% |
| 009A | 66.6% | 76.2% | 57.1% | 76.2% |
| 010A | 66.7% | 72.2% | 50.0% | 83.3% |
| 011A | 66.7% | 73.3% | 73.3% | 60.0% |
| 012A | 60.0% | 73.3% | 86.7% | 80.0% |
| 013A | 53.3% | 66.7% | 46.7% | 86.7% |
| 014A | 53.3% | 46.7% | 60.0% | 86.7% |
| 015A | 73.3% | 80.0% | 66.7% | 60.0% |
| 016A | 53.3% | 46.7% | 66.7% | 66.7% |
| Mean | 61.3% | 62.3% | 60.6% | 65.2% |
| StDev | 12.8% | 13.0% | 15.8% | 16.3% |
| Population | 63.8% | 64.6% | 61.8% | 64.2% |

Table 11 – Average MLP neural net accuracy over three runs with reshuffling train/test dataset split each run.

## 5. Discussion

We have presented a novel, lightweight approach to extracting fundamental gait features from IMU signals generated by commercially available headphones and other wearable devices, which we describe as "movement sequencing". We have also demonstrated the advantages of approaching gait analysis as one would genetic sequencing for precision medicine - i.e. to operate at the level of the individual as opposed to the population level. This work demonstrates that the proposed features extracted from the frequency domain using approaches adapted from the motion capture literature and stellar oscillation models are robust and flexible and also are sensitive enough to capture changes to an individual's normal walk in response to simulated pathology using a single head-based IMU sensor found in commercially available headphones.

**Gait Features in the Frequency Domain**
*Robust*
By leveraging the consistent oscillation that occurs during locomotion, our featuresets are less susceptible to signal interference and artifacts than traditional gait metrics. Participants were asked to complete an outdoor circuit on a university campus that was not closed to the general public, meaning they had to navigate undulating terrain and have situational awareness as they completed the test. Working in the frequency domain lowers the impact of one-off movements (e.g. a head turn to check surroundings), while emphasizing repeated irregularities, as demonstrated by results presented above. Participants were consistent within themselves across trials, with no outliers. These features do not rely on distance measurements derived from GPS, which are highly susceptible to error due to interference from environmental objects such as tall buildings and signal disruption, and they also do not rely on user-based input of anthropometrics such as height, weight, and leg length. This robustness is a strong case for the application of these features in uncontrolled environments, i.e., in the real world, and for building models that combine these featuresets with traditional gait metrics.

*Flexible*

Each coordinate direction is informative to understanding a dynamic system as complex as the human body under locomotion. As the magnitude of the motion signal varies in each direction during normal bipedal walking (e.g. raw mediolateral signals are considerably smaller in magnitude than anteroposterior and vertical signals [27]), it is desirable that the method of feature extraction is flexible enough to accommodate signal variability by sensor direction. These results indicate that the proposed featuresets are effective measurements of gait across sensor directions. Every participant had at least one feature that was significantly different between NB and B walking, with an average of 10 significant features per participant. These features are also sensitive across sensor directions; across participants, an average of 4 features were significantly different between NB and B walking in each sensor direction (acceleration X, acceleration Y, acceleration Z, and rotation Y). Therefore, these features can be used to analyze whole-body gait mechanics across coordinate directions.

*Sensitive*

Our results demonstrate that the proposed featuresets are sensitive enough to capture individual responses to pathology outside of a laboratory environment. There are significant differences ($p < 0.05$) between NB and B walking across featuresets in both our population-level pairwise testing and individual-level pairwise testing. When these features are used as model attributes in a simple MLP neural net, the model was able to label NB and B with 63.6% and 62.4% accuracy at the population level and participant level, respectively. This indicates that these featuresets are informative as standalone metrics, even without including modalities such as traditional gait metrics in the model.

**Aggregate vs Individual**

In controlled, similar systems, we expect that exposure to the same external event will produce a consistent response across each system. Human locomotion is not one such system. While the gross pattern of bipedal locomotion is consistent, the details of individual movement patterns and adaptations to external stimuli (e.g. orthopaedic braces simulating pathology, terrain changes, etc.) appear to be unique to the individual. Across our analyses, no single feature or featureset was consistently significant in all participants, which supports the prior literature that an individual's walking mechanics are unique [12]. That there was no single feature or set of features that are significant across the brace testing also suggests that individuals adapt their gait mechanics to pathology using different mechanisms. No consistent mechanism exists for adaptation in any sensor direction. This is a challenge to drawing conclusions about deviation from "normal" from aggregate data. Our population-level analyses also showed a qualitatively broader spread in principal component strengths than our individual-level analyses, further suggesting that comparing individuals to the population has limited applicability and motion adaptations are unique to the individual.

Algorithms for processing movement data are typically developed through laboratory testing using a convenience sample, which, even with large numbers of participants, may not capture the full extent of human variability. Our findings highlight how population-level analyses do not always reflect individual-level results. In our Wilcox testing, although G10 was not significantly different at the participant level in any direction, it was significantly different at the population level in acceleration Y (anteroposterior acceleration). Comparatively, while M5 was significantly different for 58.8% of participants when assessed individually, it was non-significant at the population level in any sensor direction. Our PCA results also reflect the lack of alignment between population-level and individual-level results. Some individuals' PCA results are similar to the population. For PC1 and PC2 across sensor directions participants 016A and 003A share 5 out of 16 top features and 4 out of 16 top features with the population-level PCAs, respectively, as seen in the top features highlighted in Table 6. However, the population-level results do not reflect most of the individuals' results; across sensor directions, participants shared an average of 2 top features out of 16 in PC1 and PC2.

In addition, the limited size of our datasets may have also contributed to the relative strength of individual features in individual-specific datasets when compared to the complete population-level data. In this study we collected ~15-20 minutes of data on each individual, or approximately 16MB per participant, ~270MB across the participant sample. We slice our data in order to examine each sensor direction separately, although this reduced the amount of data available for training and testing. Even with the constraints in sample size, our simple MLP was able to consistently differentiate between NB and B walking at both the population level and at the level of the individual. We returned similar model accuracy for the population sample and individual sample, even though the population sample size was considerably larger. This suggests that individual-level models would potentially outperform population-level models with larger datasets.

**Constraints & Future Work**
The aim of this pilot study was to determine whether the novel features extracted from IMU signals collected at the head could differentiate between normal and pathological walking. One of the potential limitations of this study, and of the individual-level approach more broadly, is that it relies on capturing the full range of variability in the individual's 'normal' walking to accurately identify pathological walking. This study collected data on individuals at one point in time, but future work should collect longitudinal data to better understand normal variability within individuals. Gathering more data on each individual would also expand the ability to test different deep learning model configurations as a limited sample size constrained our ability to test without overfitting the model. The point of this approach is to better characterize human movement in real world environments rather than controlled environments; however future work is needed to test these features alongside traditional gait lab data modalities.

Future work should continue to develop and test these novel gait features in both real world and laboratory conditions. As discussed above, environmental features such as gradient, terrain, and weather influence gait mechanics; therefore future models will include modalities describing environmental conditions. Using a basic neural net, we demonstrate the potential of our featuresets and individual-level approach in machine learning applications. An advantage of the MLP is that, after training, it outputs not only the categorization of new data (in our testing, labeled normal 'No Brace' walking vs 'Brace' walking), but also the importance of the features in the model. In future work this will facilitate additional steps towards interpretability. Due to the multimodal and open-ended nature of humans interacting with and navigating their world, we suggest that, similar to the rapid advances seen across fields ranging from biology to climate modeling, the development of novel deep learning models will be critical to unlocking new applications for the massive amounts of data being generated by wearable devices.

6. **Conclusion**
   Many consider walking "boring" movement, yet it is one that most people utilize repeatedly in daily life.

As such, walking is a strong system wellness check - an at-home measurement tool for overall health and well-being. If approached as one would approach genetic sequencing, i.e. from the level of the individual, this method of quantifying normal baseline movement has the potential to unlock a new realm of predictive health. With the rise of wearable technology, people now have access to large quantities of individual-level information, which provides a level of granularity that can be leveraged to bring precision medicine outside of the clinic. The benefits of at-home monitoring as a supplement to clinical care are already evident in fields such as cardiac medicine. As with all human data, the level of sensitivity required in an analysis is dependent on the question being asked. Population-level data answer questions about generalized human behavior, but, depending on the sample and the range of human variability, may not be representative of specific humans. At the same time, the wider the range of variability in a dataset, the harder it is to identify responses to condition changes that are significant at both the population-level and individual-level. This is especially challenging in complex systems such as human locomotion, where many ways to adapt gait mechanics to conditions such as pain, pathology, or environment are used by individuals. For instance, one person might respond to a strained ankle by taking shorter steps while another might limit ankle range of motion. An additional layer of complexity is manifest if the control of consistency in laboratory conditions is removed and gait is examined in the real world. In this work we demonstrate that a viable solution to these challenges is to use the individual as their own control. Since individual-level data has less variability than population-level data, we reduce the number of controlled variables, and, by examining individual-level data, we increase the sensitivity of our models, expanding opportunities for personalization outside of the clinic.

Overall, in this paper we propose a novel set of features that can be extracted from human movement using IMU data from commercially available headphones, as well as a novel approach to identifying gait-based biomarkers of pathology by targeting individual variability. We demonstrate that these features are sensitive measurements that capture nuances within an individual's gait mechanics, distinguishing one individual from another and allowing us to precisely identify subtle shifts in gait mechanics in response to pathology.